\documentstyle{mn}
\input epsf.sty

\newif\ifAMStwofonts
\AMStwofontstrue
 
\def\LE{L/L_{Edd}}
\def\Ly{Ly_{\alpha}}
 
\ifoldfss
  \ifCUPmtlplainloaded \else
    \NewTextAlphabet{textbfit} {cmbxti10} {}
    \NewTextAlphabet{textbfss} {cmssbx10} {}
    \NewMathAlphabet{mathbfit} {cmbxti10} {} 
    \NewMathAlphabet{mathbfss} {cmssbx10} {} 
  \fi
  \ifAMStwofonts
    \ifCUPmtlplainloaded \else
      \NewSymbolFont{upmath} {eurm10}
      \NewSymbolFont{AMSa} {msam10}
      \NewMathSymbol{\upi}     {0}{upmath}{19}
      \NewMathSymbol{\umu}     {0}{upmath}{16}
      \NewMathSymbol{\upartial}{0}{upmath}{40}
      \NewMathSymbol{\leqslant}{3}{AMSa}{36}
      \NewMathSymbol{\geqslant}{3}{AMSa}{3E}

    \fi
  \fi
\fi 
 
\ifnfssone
  \newmathalphabet{\mathit}
  \addtoversion{normal}{\mahit}{cmr}{m}{it}
  \addtoversion{bold}{\mathit}{cmr}{bx}{it}
  \newmathalphabet{\mathbfit} 
  \addtoversion{normal}{\mathbfit}{cmr}{bx}{it} 
  \addtoversion{bold}{\mathbfit}{cmr}{bx}{it}
  \newmathalphabet{\mathbfss} 
  \addtoversion{normal}{\mathbfss}{cmss}{bx}{n}
  \addtoversion{bold}{\mathbfss}{cmss}{bx}{n}
  \ifAMStwofonts
    \ifCUPmtlplainloaded \else
and
your
      \UseAMStwoboldmath
      \makeatletter
      \new@mathgroup\upmath@group
      \define@mathgroup\mv@normal\upmath@group{eur}{m}{n}
      \define@mathgroup\mv@bold\upmath@group{eur}{b}{n}
      \edef\UPM{\hexnumber\upmath@group}
      \new@mathgroup\amsa@group
      \define@mathgroup\mv@normal\amsa@group{msa}{m}{n}
      \define@mathgroup\mv@bold\amsa@group{msa}{m}{n}
      \edef\AMSa{\hexnumber\amsa@group}  
      \makeatother
      \mathchardef\upi="0\UPM19
      \mathchardef\umu="0\UPM16
      \mathchardef\upartial="0\UPM40
      \mathchardef\leqslant="3\AMSa36
      \mathchardef\geqslant="3\AMSa3E
    \fi
  \fi
\fi 
   
\ifnfsstwo
  \DeclareMathAlphabet{\mathbfit}{OT1}{cmr}{bx}{it}
  \SetMathAlphabet\mathbfit{bold}{OT1}{cmr}{bx}{it}
  \DeclareMathAlphabet{\mathbfss}{OT1}{cmss}{bx}{n}
  \SetMathAlphabet\mathbfss{bold}{OT1}{cmss}{bx}{n}
  \ifAMStwofonts
    \ifCUPmtlplainloaded \else
      \DeclareSymbolFont{UPM}{U}{eur}{m}{n}
      \SetSymbolFont{UPM}{bold}{U}{eur}{b}{n}
      \DeclareSymbolFont{AMSa}{U}{msa}{m}{n}
      \DeclareMathSymbol{\upi}{0}{UPM}{"19}
      \DeclareMathSymbol{\umu}{0}{UPM}{"16}
      \DeclareMathSymbol{\upartial}{0}{UPM}{"40}
      \DeclareMathSymbol{\leqslant}{3}{AMSa}{"36}
      \DeclareMathSymbol{\geqslant}{3}{AMSa}{"3E}
    \fi
  \fi  
\fi 
      
\ifCUPmtlplainloaded \else
  \ifAMStwofonts \else 
    \def\upi{\pi}
    \def\umu{\mu}
    \def\upartial{\partial}
  \fi
\fi

\title{The role of extended corona in formation of emission lines
and continuum in AGN}
\author[A. Kurpiewski, J. Kuraszkiewicz, B. Czerny]
       {A. Kurpiewski$^1$, J. Kuraszkiewicz$^2$, and B. Czerny$^2$\\
        $^1$Astronomical Observatory of Warsaw University, Aleje Ujazdowskie
4, 00-478 Warsaw, Poland\\
        $^2$N. Copernicus Astronomical Centre, Bartycka 18, 00-716 Warsaw, Poland}

\begin{document}

\maketitle

\begin{abstract}

We study a model of an accretion disc surrounded by an extended corona
of the temperature $10^7 - 10^8$ K. This corona modifies the disk spectrum 
since it redirects a significant fraction of the emission from the central 
parts towards the more distant parts of the disk. The same corona is 
indirectly the source of the broad emission lines because we expect the 
formation of cool clouds at the basis of the corona due to thermal 
instabilities. We model the number of the clouds and their motion through 
the corona adopting a few different physically sound assumptions.

Comparing the predicted optical/UV continua and emission line ratios and 
profiles with the observed values we favor a particular model of a typical 
quasar. It radiates at $\sim 0.5$ of the Eddington luminosity, the corona
surrounding the disk is additionally heated in excess of the Inverse Compton 
heating, and the broad line clouds are most probably destroyed when accelerated
vertically above the sound speed within the corona.

\end{abstract}


\section{Introduction}

There is a general agreement that the nuclear activity of galaxies
(with exception of starburst galaxies) is powered by accretion of
interstellar gas onto a massive black hole (e.g. Rees 1984). The accretion
flow most probably forms a kind of disc structure due to an excess of 
angular momentum. 
 
The evidence of such disc-like structure is observed directly in HST pictures 
of the galaxy M87 at 
a distance of several pc from the center (Harms et al. 1994) and even up to 0.009
pc in the case of the galaxy NGC 4258 due to the presence of a water maser
(Greenhill et al. 1995). 
Closer in the accretion flow is clearly two-phase. The cold gas most probably 
 forms a 
relatively flat configuration, either in a form of an accretion disc
or blobs. This material is embeded in a hot medium. 

There are several
observational arguments in favour of this general scenario.

Disc-like geometry of the cold phase explains the lack of
significant absorption in quasars and Seyfert 1 galaxies  
as well as the presence of the X-ray
spectral features in Seyfert 1 galaxies  
due to reprocessing by cool optically thick gas (i.e. the reflection 
component) covering approximately
half of the sky for the source of X-rays (e.g. Pounds et al 1990, 
Matsuoka et al. 1990; for a review, see Mushotzky, Done \& Pounds 1993). 
Additional support
for the disc-like geometry and Keplerian motion comes from determination
of the shape of the $K_{\alpha}$ line from ASCA data (Fabian et al. 1994). The
reflection component is not seen in quasars (Williams et al. 1992) 
most probably due to higher
ionization stage of the gas (e.g. \. Zycki et al. 1994, \. Zycki \& Czerny 1994).

The hot optically thin medium is required to explain the generation of the
observed hard X-ray emission extending up to a few hundreds of keV (see
Mushotzky et al. 1993). It most probably
forms a kind of corona above the cold layer of the gas 
although an  alternative view was also suggested
(X-ray emission comming from shocks formed
by gas outflowing along the symmetry axis, e.g. Henri \& Pelletier 1991).   
 
The heating mechanism of the corona, its structure and radial extension are
presently unknown.

In the innermost parts of the flow a (sometimes considerable) fraction
of the total gravitational energy of accreting gas has to be dissipated 
in the hot corona to provide both the X-ray bolometric luminosity and the
extension of the emitted spectrum into high frequencies.  Most probable
emission mechanism is  the Compton upscattering of the photons
emitted by the cool gas by hot thermal electrons.
The medium is not necessarily uniform, as we observe both the effect of
moderate Comptonization 
(modification of the high frequency tail of the big bump emission 
seen in soft X-ray band in a number of sources, e.g. Czerny \& Elvis 1987,
Wilkes \& Elvis 1987, Walter \& Fink 1993) as well as significant (but still
unsaturated) Comptonization which leads to formation of the hard X-ray 
power law.  It suggests that hard X-ray emission is perhaps produced 
in hotter, maybe magnetically driven compact active regions (e.g. Haardt,
Maraschi \& Ghisellini 1994, Stern et al. 1995) embedded, or surrounded by
still hot but cooler plasma. Stochastic nature of the X-ray variability
supports this view (Czerny \& Lehto 1996).

In the outer parts of the flow the amount of energy available is small and
the radiation emitted there does not practically contribute to the 
bolometric luminosity of the source and these regions are even less 
understood than the innermost flow. On the other hand the formation of
the Compton-heated corona seems to be inevitable if only the outer parts of
the disc are not shielded from the the radiation 
comming from innermost parts (e.g.
Begelman, McKee and Shields 1983, Begelman \& McKee 1983, 
Ostriker, McKee \& Klein 1991 - hereafter OMK, Raymond 1993). The existence
of such a corona is actually observed  in X-ray binaries (e.g. 
White \& Holt 1982, McClintock et al. 1982, Fabian \& Guilbert 
1982). The existence of such a corona in AGN may have very
significant influence on the observed spectrum.

The principal role of the corona surrounding the disc-like flow 
at a radius $\sim 0.01 - 1 $ pc is not in the 
direct dissipation of the energy but in redirecting the radiation generated
in the inner parts towards outer parts of the disc by almost elastic
scattering. The direct irradiation in the case of AGN is not efficient
unless the source of radiation is situated high above the disc surface.
The disc surface in AGN does not flare in its inner parts, 
according to the widely adopted description of the radiation pressure 
dominated disc by Shakura and Sunyaev (1973) and  in its outer 
flaring parts does not cover more than a per cent of the sky of the 
central X-ray source (e.g. Hure et al. 1994, Siemiginowska, Czerny \& 
Kostyunin 1996). 

The irradiation of the disc surface due to the scattering by the 
corona has two consequences:
(i) the irradiation modifies the optical/UV 
continuum emitted by the disc and (ii) it leads to line
formation at the basis of the corona thus contributing significantly 
to the observed Broad Emission Lines.
Both effects are the subject of our study, with the aim to confirm the
presence of the outer corona and to constrain its properties.

The plan of this paper is following. The description of the model 
of the corona and the method of
computation of the continuum emission of the disc and line intensities
and profiles from clouds forming in the disk/boundary layer  is
described in Section 2. In Section 3 we present the results for a range 
of model parameters and compare them with observations as well as we
discuss the consequences of the model. Conclusions are given in Section 4.

\section{The model of the continuum and line emission}

\subsection{Structure of the corona and location of the source of photons}

The model of the corona above the outer parts of an accretion disc ($0.01 -
1 $ pc) is based on the theory of the two-phase equilibrium studied originally
by Spitzer (1978) and further developed in the context of AGN in a number
of papers, starting from Krolik, McKee and Tarter (1981) (hereafter KMT).

The temperature of the corona is mainly constrained by the inverse Compton 
heating and cooling by the incident radiation. 
The corona is irradiated by three different radiation sources: the
high-energy central UV/X-ray source, radiation from the central source
that has been scattered in corona and lower energy radiation from
underlying viscous accretion disc. 
The central hard X-ray emission and the central thermal emission 
(for the most part UV) of the
disk are both represented as a point like
source located at a height $H_X$ along the symmetry axis.
Although in reality both emission regions are extended most of the
energy is released within  a few
gravitational radii from the black hole which is $\sim 10 ^{-4}$ or less
of
the the outer radius of a corona so the extension of the central source
can be neglected.
In our calculations we take half of the total disk luminosity as the
value of central thermal luminosity. The proportion of total X-ray 
luminosity to central thermal luminosity is discussed in Sec. 2.3.1.

Folowing KMT and OMK we denote the inverse Compton temperature of the
direct radiation (i.e. from the central source) by $T_{IC}$.
For the
spectrum of quasars used by KMT the value of $T_{IC}$ is 
$\sim 10^8$ K (KMT), but
allowing for more emission from "big blue bump" reduces $T_{IC}$ to
$\sim 10^7$ K (Mathews \& Ferland 1987, Fabian et al. 1986). The presence of
additional heating besides inverse Compton process does not modify the 
basic picture (if the heating rate
is proportional to the gas density, e.g. 
Yaqoob 1990). However, it can raise the temperature of the corona above
the inverse Compton limit, thus decoupling its value from the shape of the
incident radiation spectrum. Therefore, we can treat the maximum value of
the surface temperature of the corona, $T_C$, as a free parameter of the model. 

Including scattered and disc components of irradiation, as well as
bremsstrahlung cooling process near base of the corona, the temperature
of the corona (constant in the vertical direction) is (OMK)

  \begin{equation}
   T_{cor(r)}=\frac{T_{C}}{2}\frac{F_{dir(r)}+F_{scat(r)}+F_{visc(r)}(T_{visc(r)}/T_{C})}{F},
  \end{equation}
where
  \begin{equation}
   F_{(r)}=F_{dir(r)}+F_{scat(r)}+F_{visc(r)}
  \end{equation}
is total radiative flux at the base of the corona from all sources of
irradiation. Indices 'dir', 'scat' and 'visc' are related to direct,
scattered and disc components respectively and $T_{visc}$ is the 
temperature of radiation emitted locally from the disc ($F_{visc}$).

The fraction of the central direct radiation ($F_{dir}$) and radiation 
 scattered by the corona ($F_{scat}$) towards the disc at a radius $r$ 
is given by simple analytical formulae of OMK in the case of
sources with low ratio of the luminosity to the Eddington luminosity.

If the luminosity of the source is closer to the Eddington luminosity the
corona is optically thicker and multiple scatterings play a significant role.
In this case the numerical computations are necessary but the results are
given in the paper of Murray et al. (1994) - hereafter MCKM -  for a few 
sets of model parameters.

We use both the papers of OMK and MCKM to calculate the
direct and scattered radiative flux irradiated the disk at a given radius.
Therefore we follow the assumptions about the corona structure and the
geometry made in those papers. 

The corona extends up to 0.2 of the true maximum radius where the thermal
energy of the gas particle $\sim kT_C$ is
balanced by gravitational energy of the particle $\frac{GM\mu}{r}$
 
  \begin{equation}
  r_{C}=\frac{GM\mu}{kT_{C}}\approx\frac{10^{10}}{T_{C8}}              
  \left(\frac{M}{M_{\odot}}\right) cm,
  \end{equation}
where $T_{C8}$ is 
the maximum surface temperature of the corona expressed in units of 
$10^8$ K, $\mu$ is the mean mass per particle ($\mu=0.61m_{p}$ for fully
ionized gas of cosmic abundances) and $M$ is the mass of the
black hole.

The ionization parameter $\Xi$ (KMT) at the basis of the corona is equal
(McKee \& Begelman 1990)
  \begin{equation}
   \Xi_{b(r)}\approx 1.3T^{-3/2}_{cor8(r)} ,
   \end{equation}
assuming that at higher densities bremsstrahlung is the only atomic cooling
process. Since the ionization parameter is defined as a ratio of the incident
radiation pressure to the gas pressure, its value determines the density
at the basis of the corona (see Section 2.3.2).

\subsection{Computation of the optical/UV continuum}

\subsubsection{Heating by viscous dissipation}

Accretion disc is heated by dissipation of the gravitational energy of
the gas through viscous forces. The radiative flux which corresponds with
this energy for the disc with nonrotating black hole is given by (Page \& Thorne
1974)
  \begin{equation}
  F_{visc(r)}=\frac{3GM\dot{M}}{8\pi
  r^3_g}\cdot f_{(r)} 
  \end{equation}
  \begin{equation}
  f_{(r)}=\frac{\left[\sqrt{r}-\sqrt{3}+\sqrt{\frac{3}{8}}
  \ln\left(\frac{2-\sqrt{2}}{2+\sqrt{2}}\frac{\sqrt{r}+\sqrt{\frac{3}{2}}}
  {\sqrt{r}-\sqrt{\frac{3}{2}}}\right)\right]}{r^{\frac{5}{2}}(r-\frac{3}{2})} ,
  \end{equation}
where
  \begin{equation}
  r_g=\frac{2GM}{c^2}\approx 3.01\cdot  
  10^{-5}T_{C8}r_{C}
  \end{equation}
is the Schwarzschild radius, 
and
  \begin{equation}
  \dot{M}=\frac{L_{visc}}{\varepsilon c^{2}}
  \end{equation}
is the accretion rate. 
The efficiency $\varepsilon$ is $\sim 5.6$\% for a nonrotating black hole. 
In equations (5) and (6) the radius r is expressed in $r_g$ units.

This picture does not leave any room for the X-ray emission.  Actually, a 
fraction of energy due to accretion is dissipated in the form of X-rays. 
However, reliable predictions of that fraction as a function of radius are
not available (see e.g. Witt et al. 1996). On the other hand the efficiency
of accretion is most probably higher than adopted as the black hole may
well be rotating, increasing the bolometric luminosity easily by a factor few.
Therefore we describe the accretion disc emission as above but we allow
that the total luminosity of the central source $L$ may be higher than the
disc luminosity $L_{visc}$ due to the contribution from unaccounted X-ray 
emission.

\subsubsection{Direct radiative heating of the disc}

The additional radiation flux heating the disc is the flux of radiation
from the central UV/X-ray source with luminosity L
  \begin{equation} 
  F_{dir(r)}=(1-A)\frac{L}{4\pi({r}^2+H^2_X)}
   f_{dir(r)}cos{\theta}_{dir(r)}
  \end{equation}
where A is the albedo of the disc, ${\theta}_{dir(r)}$ is the angle between
the incident ray and the normal to the disc at $r$, and 
  \begin{equation}
   f=\frac{F}{L/4\pi r^2}
  \end{equation}
is used by OMK and MCKM dimensionless factor, which is
the ratio of the radiation flux at the base of the corona to the
unattenuated flux from the central source (in this definition in the
denominator the component $H^2_X$ is neglected). To determine the factor
$f_{dir}$ we use the results of OMK or MCKM papers as we mentioned in
Sec. 2.1.
We assume A=0.5 as an appropriate value for quasars because the disc surface
is partially ionized (see e.g. \. Zycki et al. 1994).

\subsubsection{Heating by corona}

To describe the effect of irradiation by photons from the central source
scattered towards the disc surface by extended corona we also use the method 
described in Sec. 2.1 to determine the scattered radiation flux

 \begin{equation}
  F_{scat(r)}=(1-A)\frac{L}{4\pi r^2}
  f_{scat(r)} \langle cos{\theta}_{scat} \rangle ,
  \end{equation}
where $\langle {\theta}_{scat} \rangle $ is the angle between the 
direction from the scattering point and the normal to the disc at $r$, 
averaged over the whole volume of the corona. We determine the factor 
$f_{scat(r)} \langle cos{\theta}_{scat} \rangle$ from OMK or MCKM papers.
We adopt the same albedo as for the direct flux.

\subsubsection{Computation of the spectrum}

The effective temperature of the disc photosphere is given by 
  \begin{equation}
  T_{eff(r)}=\left(\frac{F_{visc(r)}+F_{dir(r)}+F_{scat(r)}}{\sigma}\right)^{\frac{1}{4}}
   .
  \end{equation}
Assuming the black body emission from the disc, one can calculate the shape 
of continuum as follows
  \begin{equation}
  f_{\nu}=2\pi \intop_{3r_g}^{r_{max}}rB_{\nu}[T_{eff(r)}]dr,
  \end{equation}
where $B_{\nu}$ is the Planck function and $r_{max}$ is the outer radius
of the disc (in calculations we assume $r_{max}=r_{C}$).

We neglect the modification of the disc spectrum due to electron scattering. 
Although such effects were discussed by a number of authors (e.g. 
Czerny \& Elvis 1987, Ross \& Fabian 1993, Shimura \& Takahara 1995)
these corrections strongly depend on the accuracy of the description of atomic
processes as well as assumptions on the disc viscosity.

\subsection{The spectrum of the central source and the emission lines}

\subsubsection{Incident radiation spectrum}

The computations of the continuum given in Sect. 2.2 cover only
the optical/UV band as only the knowledge of the total X-ray luminosity, but
not the shape of the primary X-ray emission, was required to compute the 
thermal emission of the disc dominating in this spectral band. On the other
hand the computations of the strength of the emission lines require the
determination of the spectrum in the EUV and X-ray band as well. 

Therefore, for the purpose of calculating emission lines we use the spectral
shapes which approximate well the observed overall spectra of Seyfert galaxies
and quasars.

The shape of the big blue bump is parametrized in a similar way as by
Mathews \& Ferland (1987). However, we adjusted parametrization in agreement
with the present observational data and we adopted interpolation to give
the ratio of the bolometric luminosities of the big blue bump to X-rays
equal $\sim 1$ for Seyfert galaxies and $\sim 10$ for quasars. The details
of this parametrization are given in Appendix. The inverse Compton
temperatures for these two spectra are $5.8\cdot 10^{7}$ K for Seyfert
galaxies and $1.1\cdot 10^{7}$ K for quasars.

\subsubsection{Local line emissivity}

We assume that the high ionization emission lines come from the clouds which 
form in the narrow intermediate zone between an accretion disc and a hot
corona. Thermal
instability of irradiated gas at intermediate temperatures causes discontinuous
transition between a disc and a corona (Begelman, McKee \& Shields 1983), if
only radiative processes are taken into account. In realistic situation, when
some level of a turbulence is present in the medium, we may expect the
spontaneous formation of the cool clumps embedded in the hot coronal plasma
in a relatively narrow transitory zone (R\' o\. za\' nska \& Czerny 1996).

The column density of the clouds is not determined precisely but the estimates
based of the relative efficiency of condution and thermal processes give
values of the same order as values expected for the clouds forming the
Broad Line Region. Therefore we assume for all the clouds the column density
equal $10^{23}$ cm$^{-2}$.

Cloud parameters (density $n_{cl}$ and temperature $T_{cl}$) at a given 
radius are determined 
by two requirements. This first condition is the pressure equilibrium 
with the hot medium at the basis of the corona. The second condition, in the
case of optically very thin plasma, should be the settlement on the lower
stable branch at the $\Xi - T $ curve (KMT). However, this branch is not applicable 
for media of higher optical depth. Since we do not expect clouds to cool
below the local effective temperature of the disk surface, we assume their
temperature to be at that value, $T_{cl}=T_{eff}$. 
Therefore we adopt the condition
 
  \begin{equation}
  n_{cl(r)}=n_{d(r)}=n_{b(r)}\frac{T_{cor(r)}}{T_{eff(r)}}=
  n_{b(r)}\frac{T_{cor(r)}}{T_{cl(r)}}
  \end{equation}
where $n_{b(r)}$ is the density at the basis of the corona determined from
the value of the ionization parameter $\Xi_b$ and $T_{cor(r)}$ and 
$T_{eff(r)}$ are given by eq. (1) and (12).

We discuss the kinematics of the clouds in Sect. 2.3.3 since it is essential
for computation of the line profiles. However, cloud motion
influences the line intensities as well.

Since the clouds are blown out radiatively from the formation region they
do not form a flat layer on the top of a disc but they are fully exposed
to the incident radiation flux so the inclination angle of the direct incident
flux does not have to be included through the cosine factor, as it is the case
for the disk surface. 

We calculate the emissivity of several emission lines (see Table 5 for the 
list) using the photoionization code CLOUDY. The calculations are made
for a grid of radii resulting from the used range of $n_{cl}$: $10^9 -
10^{13} cm^{-3}$ (the same for all models).
For each radius separately we calculate
the contribution to emission lines assuming the local value of the
total heating flux $F_{(r)}$ given by eq. (2),(5),(6),(9),(11), density 
$n_{d(r)}$ from eq. (14) and fixing the column density $N_{H}=10^{23}$ cm$^{-2}$. 

\subsubsection{Radial dependence of number of clouds}

Clouds forming in the transition layer between the disk and the corona does
not necessarily cover the disk surface uniformly. Local number of clouds
give weight to the local emissivity thus influencing both the line ratios
and line profiles. The number of clouds, $ N _{(r)}$,  
existing at a given radius depends
both on the cloud formation rate, $\dot N_{(r)}$, and expected life time of 
a cloud, $t_{(r)}$
\begin{equation}
   N _{(r)}=\dot N_{(r)} t_{(r)}.
\end{equation}

As the detailed process of cloud formation and destruction is not well
understood we discuss a few representative cases based on available estimates.

We consider two cases of the cloud formation process. In the first case we
assume that only one cloud can form at a given moment and a given radius
and the formation time is given by the characteristic isobaric cooling time,
$\tau_{(r)}$ (McKee \& Begelman 1990)
\begin{equation}
 {\rm case (I)}~~~~~ \dot N_{(r)}\sim \frac{1}{\tau_{(r)}}\sim n_{b(r)} .
\end{equation}
In the second case we assume that the clouds form in the entire 
instability zone therefore the number of the clouds forming at the same time
is related to ratio of the zone geometrical thickness $\Delta Z$ , which
is of order of the Field lenght (e.g. R\'o\.za\'nska \& Czerny 1996)
to the size of the cloud, $r_{cl(r)}=N_{H}/n_{cl(r)}$ 
\begin{equation}
  {\rm case (II)}~~~~~ \dot N_{(r)} \sim \frac{\Delta Z_{(r)}}{r_{cl(r)} \tau_{(r)}}
   \sim n_{cl(r)} .
\end{equation}

We describe the destruction process using three different approaches.
In the case (i) we assume that the clouds survive only within the instability 
zone. As they move through the zone upwards 
under the influence of the radiation from the disk surface their life time
is given by the travel time through the zone $\Delta Z$. We assume that
the radiation acceleration is constant. It gives us the relation
\begin{equation}
  {\rm case (i)}~~~~~  t_{(r)}  \sim \frac{1}{\sqrt{n_{b(r)}}} .
\end{equation}
  
In the case (ii) we assume that clouds survive even outside the instability 
zone but undergo the destruction due to conduction, evaporating into 
surrounding hot corona. The timescale of such a process
is given by McKee \& Begelman (1990) 
\begin{equation}
  {\rm case (ii)}~~~~~  t_{(r)}  \sim \frac{1}{n_{b(r)}} .
\end{equation}

In the case (iii) we assume that clouds are accelerated so efficiently by the
radiation pressure that they reach soon the velocities exceeding the local 
sound velocity in the corona. Reaching this terminal velocity of $\sim 
2000$ km/s
they are destroyed by dynamical instabilities. Since we assume that
the radiation acceleration of the clouds is independent from the
disk radius  the time needed to reach
this terminal velocity is also constant if small variations of the corona
temperature with the radius are ignored. In that case
\begin{equation}
  {\rm case (iii)}~~~~~  t_{(r)}  = const. 
\end{equation}
We summarize the six models of the number of clouds $N_{(r)}$  in Table 1.

  \begin{table}
  \caption{The models of radial dependence of number of clouds.}
  \begin{center}
  \begin{tabular}{|l|l|l|}
   \hline
    Model & Cases from  & The radial \\
    \ & Sec. 2.3.3 & dependence \\
   \hline
    a & (I)(i) & $\sqrt{n_{b(r)}}$ \\
    b & (I)(ii) & const. \\
    c & (I)(iii) & $n_{b(r)}$ \\
    d & (II)(i) & $\sqrt{n_{b(r)}}/T_{cl(r)}$ \\
    e & (II)(ii) & $T_{cl(r)}$ \\
    f & (II)(iii) & $n_{cl(r)}$ \\
   \hline
  \end{tabular}
  \end{center}
  \end{table}

\subsubsection{Line profiles}

We assume that the velocity field of the emitting clouds consists of the
Keplerian orbital motion and the outflow perpendicular to the disc
surface.

In our computations of the orbital motion we follow the method of 
Chen and Halpern
(1989), including the relativistic effects. Their procedure requires the
specific intensity from the disc surface 
  \begin{equation}
  I_{(r ,{\nu}_e)}=\frac{1}{4\pi}{\epsilon}_{(r)}\cdot \frac
  {e^{-\frac{{({\nu}_e-{\nu}_0)}^2}{2{\sigma}^2}}}{(2\pi )^{1/2}\sigma }
  \ [ergs s^{-1}{cm}^{-2}{sr}^{-1}Hz^{-1}]
  \end{equation}
where ${\epsilon}_{(r)} $ is the local emissivity of the clouds (see
Sec. 2.3.2) and the exponential factor is
related to local broadening of the emission line. The broadening (due to
electron scattering or turbulence e.g.) is characterized by quantity
$\sigma$, and ${\nu}_e$, ${\nu}_0$ are the emitted and rest frequencies.
After Chen and Halpern (1989) we use in our calculations
$\sigma$/${\nu}_0$=0.03. 

The vertical motion is computed assuming that the acceleration of the clouds
is constant, i.e. the velocity increases linearly with the distance from
the disc surface. Maximum velocity is constrained by the cloud destruction. 

Since the line profiles are not strongly dependent on the detailed description
of the vertical motion as the orbital velocity is usually much higher than
the vertical one we do not distinguish between the three cases introduced in
Section 2.3.3 and in all cases we assume that the clouds move under the
influence of constant radiative acceleration and reach the same terminal
velocity equal 2000 km/s, independently from the disk radius, as expected in
case (iii).

\begin{figure*}
\leavevmode
\epsfysize = 110 mm \epsfbox[20 380 560 770]{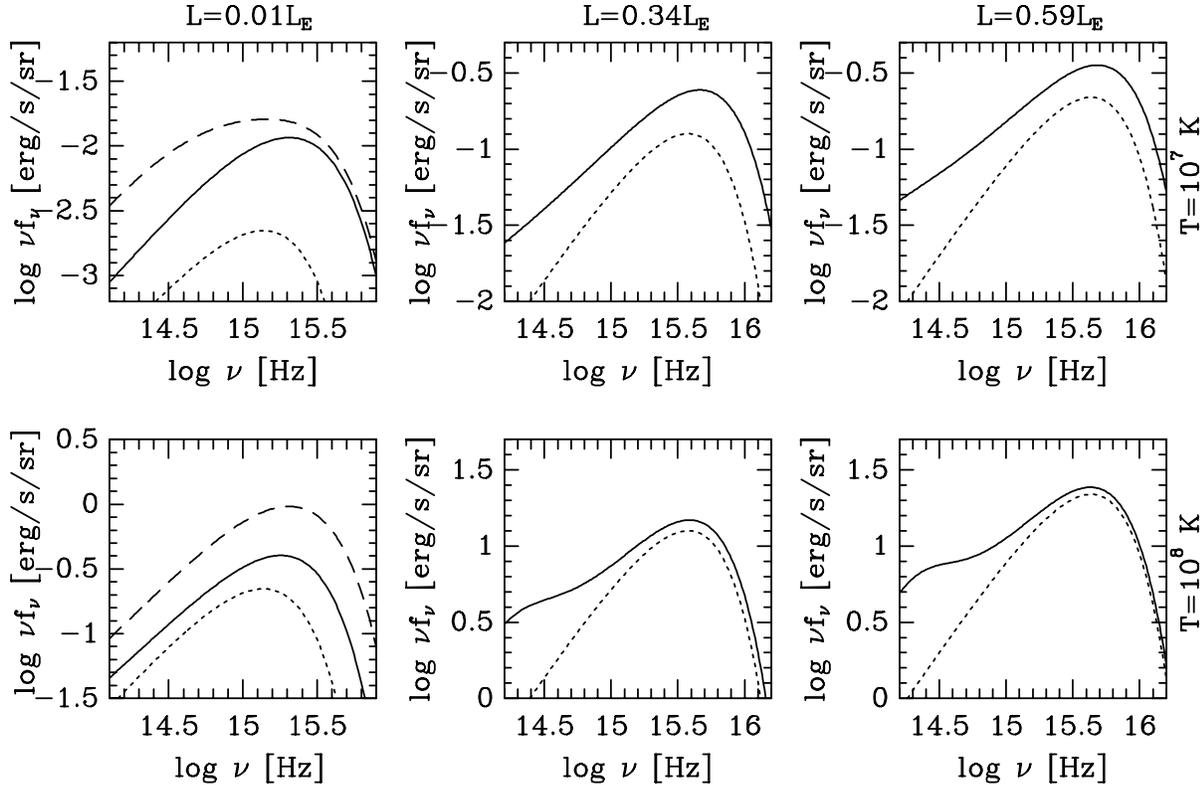}
\caption{The spectra of irradiated accretion disks parametrized
by the $\LE$ ratio and the corona temperature. The dotted curves show 
the spectra of non-irradiated discs and the dashed ones illustrate the
cases with $H_X=33.2r_g$.}
\end{figure*}

\section{Results and discussion}

We calculate the local radiation flux at the disc surface for
the following parameters 
  \[
  M=10^8M_{\odot}
  \]
  \[
  \varepsilon =0.056
  \]
  \[
  T_{C}=10^8 K; 10^7 K
  \]
  \[
  L=0.01 L_{E}; 0.34 L_E; 0.59 L_E
  \]
  \[
  H_X=3.32 r_g; 33.2 r_g.
  \]

The second value of the height could only be used in the calculations of the 
low luminosity cases as the numerical solutions of the high luminosity 
corona structure (MCKM) are available only for the first value.
In agreement with generally accepted trend that low ratios of luminosity
to the Eddington luminosity are appropriate for Seyfert galaxies and
luminosities closer to the Eddington limit are appropriate for quasars,
we use Seyfert galaxies spectrum for $L=0.01 L_{E}$ and quasars spectrum
for $L=0.34 L_E, 0.59 L_E$ (see Appendix).

\subsection{IR/optical/UV continuum}
 
We calculate the accretion disk spectra taking into account the  direct
irradiation by the central source as well as the irradiation by the flux
scattered in the corona, as described in Sect. 2.2. 
Figure 1 illustrates our results in $log\nu f_{\nu}-log\nu$ form. 

In the case of low $\LE$ ratio, the corona is never strong and the direct
irradiation dominates, independently from the corona temperature. Therefore
the adopted location of the irradiating source is of significant importance.
No corona influence is seen when the temperature is lower, some redistribution
of the flux towards the outer part of the disk is present when the temperature
is higher. 

In the case of high $\LE$ ratio, the effect of the corona is essential and 
direct irradiation negligible which means that the location of the central
source is not essential any longer (unless $H_X$ would be very high indeed).
The effect of irradiation is significant even for lower corona temperature
and it is particularly strong for high temperature, leading to significant 
enhancement of spectra in IR/optical band.

  \begin{table}
  \caption{Spectral indices in the selected ranges of continuum.}
  \begin{center}
  \begin{tabular}{|l|c|c|c|}
   \hline
   \hline
  Model & $\alpha_{opt}$ & $\alpha_{UV-1}$ & $\alpha_{UV-2}$ \\
   \hline
   \hline
  $L=0.01L_E;T=10^7 K;$ &  $-$ 0.5 & $-$ 1.1 & $-$ 1.22 \\
  \multicolumn{4} {l} {$H_X=33.2r_g$} \\
  \hline
  $L=0.01L_E;T=10^7 K;$ &  0.01 & $-$ 0.74 & $-$ 0.94 \\
  \multicolumn{4} {l} {$H_X=3.32r_g$} \\
  \hline
  $L=0.01L_E;T=10^8 K;$ &  $-$ 0.06 & $-$ 0.75 & $-$ 0.94 \\
  \multicolumn{4} {l} {$H_X=33.2r_g$} \\
  \hline
  $L=0.01L_E;T=10^8 K;$ & $-$ 0.1 & $-$ 0.9 & $-$ 1.14 \\
  \multicolumn{4} {l} {$H_X=3.32r_g$} \\
  \hline
  $L=0.34L_E;T=10^7 K$ & $-$ 0.19 & $-$ 0.27 & $-$ 0.32 \\
  \hline
  $L=0.34L_E;T=10^8 K$ & $-$ 0.55 & $-$ 0.36 & $-$ 0.4 \\
  \hline
  $L=0.59L_E;T=10^7 K$ & $-$ 0.32 & $-$ 0.3 & $-$ 0.34 \\
  \hline
  $L=0.59L_E;T=10^8 K$ & $-$ 0.66 & $-$ 0.32 & $-$ 0.35 \\
   \hline
   \hline
  \end{tabular}
  \end{center}
  \end{table}

In order to compare the derived spectra and observed continuum 
shapes of accretion disc
we calculate the spectral indices $\alpha$ ($f_{\nu}\sim
\nu^{\alpha}$) in optical range ($\nu \sim 10^{14.5}\div 10^{15}$ Hz)
and in the following pieces of UV range $\nu \sim 10^{15.12}\div
10^{15.32}$ Hz (UV-1) and $\nu \sim 10^{15.22}\div 10^{15.36}$ Hz
(UV-2). They are presented in Table 2. The mean values of
spectral index for these three ranges obtained from observations by
various groups for various samples of quasars we show in Table 3. 
It could be supplemented by the UV slope (between 15.13 and 15.45) of
the composite radio quiet quasar spectrum given by Zheng et.~al (1996) 
equal -0.86, but their formal error does not reflect the dispersion of the
contributing spectra. 
Unfortunately, equaly reliable data for Seyfert galaxies are not available
since in the case of weaker active galactic nuclei the determination of the
spectral slope is complicated by strong contamination of the spectra by
circumnuclear starlight.

Comparing directly the predicted spectral slopes of the high luminosity models 
(roughly adjusted to quasar 
luminosities) with the observed values we conclude that they roughly correspond
to the data, taking into account large errors. However, the agreement with 
mean values is far from perfect since the UV slopes are too flat. This is
clearly caused by adopting the value of the black hole mass equal 
$10^8 M_{\odot}$, actually too low by a factor 3 to 10. Higher values of
the mass would give flatter optical spectra and steeper UV spectra due to the
the decrease of the disk temperature for given $\LE$ ratio. 
It might therefore slightly favor the models with higher corona temperature.
Unfortunately,
the available corona models are only for $10^8 M_{\odot}$ (see Sect. 2.1) so
we cannot support this conclusion quantitatively.

  \begin{table}
  \caption{The mean values with standard deviations of spectral index
           obtained from observations.}
  \begin{center}
  \begin{tabular}{|c|c|c|}
   \hline
   \hline
  $\alpha_{opt}$  &
  $\alpha_{UV-1}$  &
  $\alpha_{UV-2}$  \\
  Neugebauer et al. & Baldwin et al. & Francis et al. \\
   1987            &  1989 & 1992   \\
   \hline
  \  & \  & \  \\
  $-$0.4$\mp$0.4  &  $-$0.91$\mp$0.34  &  $-$0.67$\mp$0.5 \\     
  \  & \  & \  \\
   \hline
   \hline
  \end{tabular}
  \end{center}
  \medskip
   in Neugebauer et al. paper the emission in '\it small blue bump \rm '
   is subtracted.
  \end{table}

\subsection{Emission lines}

\begin{figure}
\epsfxsize = 80 mm \epsfysize = 75 mm \epsfbox[50 380 480 750]{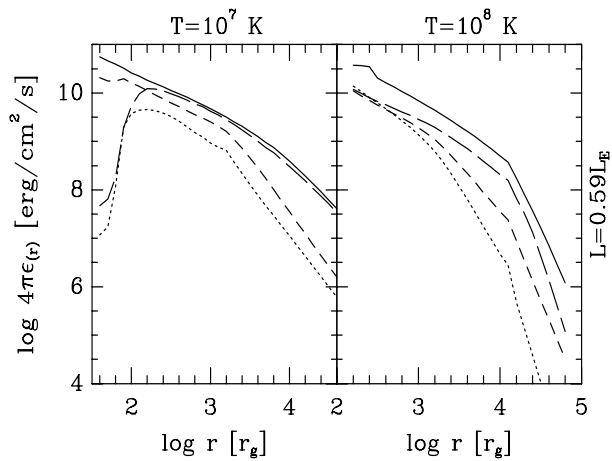}
\caption{The examples of the radial distribution of the local line emissivity
of four strongest lines as defined in Sect. 2.3.2 for $\LE$ =0.59 and the corona
temperature $10^7$ K (left panel) and $10^8$ K (right panel). The solid,
long-dashed, short-dashed and dotted curves represent $\Ly$, CIV, HeII
and NV lines respectively.}
\end{figure}

We can easily see that the emission of the broad emission lines by clouds
formed close to the basis of the hot accretion disk corona is a reasonable
assumption.
The typical widths of
lines reaching FWHM$\approx$2000 - 10000 km/s correspond to Keplerian
velocities over the radii range
\mbox{$1.3\cdot 10^{16} cm < r < 3.4 \cdot 10^{17} cm$} in the case of an 
accretion disc around $10^8 M_{\odot}$ black hole. It is the region covered
by the corona and the cloud emissivity is large in this range (see Fig. 2).
Also the line ratios should be reasonable as the cloud formation is based on
special requirements as for the ionization parameter.

Detailed predictions of the model can support this scenario and allow to
put some constraints on the $\LE$ ratio, corona temperature and the radial
distribution of clouds. 

\subsubsection{Line profiles}

  \begin{table*}
  \begin{minipage}{150mm}
  \caption{FWHMs in km/s of $Ly_{\alpha}$ and CIV lines profiles}
  \begin{tabular}{@{}lcccccccc}
    \hline
    \hline
     Model &
     \multicolumn{2}{c}{$i=0^{\circ}$} &
     \multicolumn{2}{c}{$i=30^{\circ}$} &
     \multicolumn{2}{c}{$i=60^{\circ}$} &
     \multicolumn{2}{c}{$i=80^{\circ}$} \\
     \  & $Ly_{\alpha}$ & CIV & $Ly_{\alpha}$ & CIV & $Ly_{\alpha}$ &
        CIV & $Ly_{\alpha}$ & CIV \\
    \hline
    \multicolumn{9} {c} {Model a} \\
    \hline
     $L=0.34L_E;T=10^7 K$ & 3130 & 3240 & 7070 & 7590 & 10940 & 12080
      & 12450 & 13080 \\
     $L=0.34L_E;T=10^8 K$ & 2840 & 2760 & 5890 & 6700 & 9020 & 10800
      & 9950 & 12150 \\
     $L=0.59L_E;T=10^7 K$ & 2980 & 2690 & 5750 & 5750 & 8880 & 9060
      & 9760 & 10300 \\
     $L=0.59L_E;T=10^8 K$ & 2690 & 2690 & 5194 & 4900 & 7710 & 7400
      & 8440 & 8070 \\
    \hline
    \multicolumn{9} {c} {Model b} \\
    \hline
     $L=0.01L_E;T=10^8 K;H_X=33.2r_g$ & 2700 & 2700 & 3980 & 4090 & 5670
      & 5780 & 6220 & 6340 \\
     $L=0.01L_E;T=10^8 K;H_X=3.32r_g$ & 2800 & 2800 & 4460 & 4720 & 6670
      & 7040 & 7150 & 7900 \\
    \hline
     $L=0.34L_E;T=10^7 K$ & 2620 & 2610 & 3430 & 3390 & 4680 & 4670 & 
      5200 & 5270 \\
     $L=0.34L_E;T=10^8 K$ & 2620 & 2650 & 3510 & 3790 & 5450 & 5670 &
      6010 & 6340 \\
     $L=0.59L_E;T=10^7 K$ & 2620 & 2690 & 3200 & 3280 & 4346 & 4462 &
      5000 & 5100 \\
     $L=0.59L_E;T=10^8 K$ & 2600 & 2610 & 3790 & 3790 & 5670 & 5780 &
      6300 & 6450 \\
    \hline
    \multicolumn{9} {c} {Model f} \\
    \hline
     $L=0.34L_E;T=10^7 K$ & 5120 & 2910 & 16690 & 11860 & $>$ 20000
      & $\sim$ 20000 & ... & ... \\
     $L=0.34L_E;T=10^8 K$ & 2870 & 2870 & 9320 & 9770 & 15360 & 16020
      & 17240 & 17970 \\
     $L=0.59L_E;T=10^7 K$ & 4050 & 2840 & 11050 & 9360 & 16830 & 15210
      & 19200 & 16700 \\
     $L=0.59L_E;T=10^8 K$ & 2730 & 2730 & 6810 & 6260 & 10610 & 9500
      & 11790 & 10550 \\
    \hline
    \multicolumn{9} {c} {Observations} \\
    \hline
     Brotherton et al. (1994)ALS &
     \multicolumn{4}{c}{$Ly_{\alpha}$: 8000 $\pm$ 600} &
     \multicolumn{4}{c}{CIV: 6800 $\pm$ 300} \\
    \hline
     Baldwin et al. (1989)BQS & \multicolumn{8} {c} {CIV: 2500 $\div$
      8000; mean=5680 $\pm$ 390} \\
    \hline
     Wills et al. (1993) & \multicolumn{8} {c} {CIV: 1970 $\div$
      10400; mean=4870 $\pm$ 180} \\
    \hline
    \hline
  \end{tabular}
  \end{minipage}
  \end{table*}

We present the calculated profiles for $C_{IV}$ and Ly$\alpha$ lines
only, because these lines are statistical investigated best of all.  
 
Line profiles determined by the model depend on the disk/corona model and
on the adopted radial distribution of the number of clouds, reflecting
various assumption about their formation and destruction. 

An example of the radial distribution of the local emissivity of several lines
is shown in Fig. 2. It is further combined with the radial distribution of 
the number of clouds. In the case of model (b) (see Table 1) this distribution
is preserved whilst in other models it includes additional local weight 
and either outer or inner parts of the distribution are enhanced.
 
We calculate the profiles for four values of the inclination angle of
accretion disc: 0$^o$, 30$^o$, 60$^o$ and 80$^o$. 
According to the unified scheme of
AGN supported by a number of strong observational data (e.g. 
Antonucci 1993) objects viewed at large inclination angles are obscured 
by molecular/dusty torus and are not identified as quasars so the mean 
inclination angle is expected to be smaller than 60$^o$.

We compare the profiles with the mean values and observed ranges of the
these two line widths. 

The first sample are exceptionally
well studied line profiles of quasars by Brotherton et al. (1994).
These autors analysed two quasar samples (a small sample with
 high quality UV spectra and the other from Large Bright Quasar Survay; 
hereafter ALS and LBQS) 
and  were able to decompose 
the contribution to broad lines from Intermediate Line Region 
(distant and most probably spherical) and Very Broad Line Region. 
Here we use the ALS data.

  \begin{table*}
  \begin{minipage}{180mm}
  \caption{Line Intensity Ratios}
  \begin{tabular}{@{}lccccccccccc}
    \hline
    \hline
     Model & $\Ly$ & NV & CII & SiIV & OIV] & CIV & HeII & OIII] &
     AlIII & SiIII] & CIII] \\
     \ & 1216 & 1240 & 1335 & 1397 & 1402 & 1549 & 1640 & 1663 & 1859 &
   
      1892 & 1909 \\
    \hline
    \multicolumn{12} {c} {Model a} \\
    \hline
     $L=0.01L_E;T=10^8 K;H_X=33.2r_g$ & 80 & 19 & 0 & 16 & 8 & 100 & 18
      & 4 & 2 & 2 & 3 \\
     $L=0.01L_E;T=10^8 K;H_X=3.32r_g$ & 92 & 23 & 0 & 10 & 10 & 100 & 15
      & 5 & 1 & 2 & 5 \\
    \hline
     $L=0.34L_E;T=10^7 K$ & 114 & 15 & 0 & 0 & 4 & 100 & 76 & 0 & 0 & 0
      & 0 \\
     $L=0.34L_E;T=10^8 K$ & 239 & 33 & 0 & 19 & 8 & 100 & 45 & 3 & 1 &
      3 & 3 \\
     $L=0.59L_E;T=10^7 K$ & 144 & 20 & 0 & 1 & 1 & 100 & 58 & 2 & 0 & 0
      & 1 \\
     $L=0.59L_E;T=10^8 K$ & 239 & 25 & 0 & 19 & 7 & 100 & 42 & 2 & 1 &
      3 & 2 \\
    \hline
    \multicolumn{12} {c} {Model b} \\
    \hline
     $L=0.01L_E;T=10^8 K;H_X=33.2r_g$ & 83 & 4 & 0 & 7 & 3 & 100 & 6 &
     7 & 1 & 3 & 14 \\
     $L=0.01L_E;T=10^8 K;H_X=3.32r_g$ & 84 & 6 & 0 & 6 & 4 & 100 & 6 &
     7 & 1 & 2 & 14 \\
    \hline
     $L=0.34L_E;T=10^7 K$ & 132 & 5 & 0 & 1 & 2 & 100 & 24 & 1 & 0 & 0
     & 0 \\
     $L=0.34L_E;T=10^8 K$ & 276 & 8 & 1 & 14 & 5 & 100 & 21 & 8 & 1 & 10
     & 14 \\
     $L=0.59L_E;T=10^7 K$ & 125 & 6 & 0 & 1 & 3 & 100 & 12 & 2 & 0 & 0 &
     1 \\
     $L=0.59L_E;T=10^8 K$ & 226 & 7 & 0 & 13 & 4 & 100 & 20 & 3 & 1 &
     8 & 5 \\
    \hline
    \multicolumn{12} {c} {Model f} \\
    \hline  
     $L=0.34L_E;T=10^7 K$ & 210 & 34 & 0 & 0 & 3 & 100 & 144 & 0 & 0 &
      0 & 0 \\ 
     $L=0.34L_E;T=10^8 K$ & 234 & 46 & 0 & 21 & 9 & 100 & 55 & 1 & 2 &
      2 & 1 \\
     $L=0.59L_E;T=10^7 K$ & 166 & 25 & 0 & 1 & 1 & 100 & 83 & 1 & 0 &
      0 & 0 \\  
     $L=0.59L_E;T=10^8 K$ & 250 & 37 & 0 & 22 & 9 & 100 & 54 & 2 & 2 &
      2 & 1 \\
    \hline
    \multicolumn{12} {c} {Observations} \\
    \hline
     Brotherton et al. (1994)ALS & 219 & 111 & 6 & \multicolumn{2} {c}
     {42} & 100 & 31 & 0 & 17 & $<$5 & 41 \\
     Brotherton et al. (1994)LBQS & 154 & 137 & 6 & \multicolumn{2} {c}
     {55} & 100 & 40 & 0 & 22 &  ...   & 59 \\
    \hline
     Baldwin et al. (1989)BQS & 236 & 106 & ... & ... & ... & 100 &    
     \multicolumn{2} {c} {30} & 16 & ... & 43 \\
    \hline
     Francis et al. (1991)LBQS & \multicolumn{2} {c} {159} & 4 &
      \multicolumn{2} {c} {30} & 100 & \multicolumn{2} {c} {29} &
      \multicolumn{3} {c} {46(AlIII + CIII])} \\
    \hline
     Zheng et al. (1996) & 192 & 27 & 1 & \multicolumn{2} {c} {16} &
      100 & 7 & 5 & 6 & 5 & 23 \\
    \hline
    \hline
  \end{tabular}
  \end{minipage}
  \end{table*}

The second sample (Bright Quasar Sample, BQS) is from Baldwin et al.
(1989)
and the third one quasars from Wills et al. (1993) which are not
decomposed
into ILR and VBLR, as above. In our model we expect the contribution
from
broad range of radii and such a decomposition might not be necessary.
  
In Table 4 we give the observational constraits and the results from
our most promising models, i.e. case (a), (b) and (f).
  
Radial cloud distribution (c) gives unreasonably broad profiles for both
lines
since they enhance too much the emission from inner radii (the number of
clouds decreases with radius far too fast). Models from families (d) and
(e)
gives exactly the opposite trend, leading to unreasonably narrow
profiles.
     
Model (b) also gives too narrow profiles, although not as narrow as case
(c).
Slight decrease of the cloud number with radius is clearly favored.
    
Second and third among the class (a) models well represent the CIV
distribution and require the dusty torus to cover inclination angles
above
$\sim 60^o$. However,  these solutions do not satisfy the
observational requirement that the $\Ly$ is broader than the CIV.
     
We conclude that the most favorable representation is given by model
(f),
$\LE =0.59$ and the corona temperature $10^8$ K, with the torus
shielding
the view above $\sim 60 - 70^o$. This last quantity is very sensitive to
the corona temperature. We find the solution with low temperature
equally
satisfactory if the opening angle of the torus is as small as $30^o$.

\subsubsection{Line intensities}

We compute line ratios for three models of the radial cloud distribution
which were most promissing from the point of view of $\Ly$ and CIV profiles
varying also other model parameters. The results are presented in Table 5. 

We compare them with the line ratios for quasars determined by Brotherton
et al. (1994) for ALS and LBQS samples, Baldwin et al. (1989), Francis et
al. (1991) and Zheng et al. (1996).  
 
The $\Ly$ to CIV ratio predicted by models with the luminosity close to the
Eddington luminosity and the mean quasar spectrum fall reasonably close to 
the observed values, 
although many models with low corona temperature tend to underproduce
this ratio whilst high temperture corona models sometimes overproduce it.

The dispersion between different observational results is considerable but
we see that low corona temperature, consistent with the inverse Compton
temperature of the quasar spectra, is allowed only in case (f) and $\LE$ ratio
equal 0.34; all other models require higher temperature to produce 
the required amount of $\Ly$.

Model (f), favored by the study of the line profiles, 
with the corona temperature somewhat below $10^8$ K seems quite
attractive if compared, for example, with Zheng et al. (1996).
It could reproduce well the $\Ly$ to CIV ratio as well as
the amount of NV and (perhaps) SiIV. However, it overproduces HeII line,
although this discrepancy is not so strong if other samples are considered.
It also underproduce some other weak lines like AlIII, SiIII] and CIII].

However, it is possible that there is a contribution to the line emission
from collisional excitation. In the case of clouds moving across the corona
it is only natural to expect the formation of the shock at the cloud front
and therefore some additional ionization. This effect was neglected in out 
study.

\bigskip

\section{Conclusions}

Presented results support the following view of quasars.

Quasars are radiating at $\LE$ ratio $\sim 0.5$, accretion disk in these 
objects is surrounded by a hot corona with temperature higher than the inverse
Compton temperature. Broad emission lines observed in their spectra come from
clouds which form continuously at the basis of the corona due to thermal 
instabilities and are being blown out by the radiation pressure untill they 
are destroyed when they reach supersonic velocities, $\sim 2000 $km/s.

However, further studies are necessary to confirm this picture. For example,
extension of the computations towards higher values of the mass of
the black hole and the inclusion of the contribution from collisionally heated
sides of moving clouds may be essential.

\bigskip

\section*{Acknowledgements}

We thank Gary Ferland for providing us with his photoionization code
CLOUDY version 80.07. We are gratefull to Agata R\'o\.za\'nska for
many helpful discussions.
This work was supported in part 
by grants 2P03D00410 and 2P30D02008 of the Polish State Committee for 
Scientific Research. 

\bigskip

\section*{Appendix}

We parametrize the continua emitted by the central regions of the
accretion flow by adopting fixed values of the energy index $\alpha$ 
in a number of
energy bands given in $log\nu$. In the case of Sefert galaxies we 
assume 10.5 - 14.83 ($\alpha=2.5$),14.83 - 15.76 (-0.5), 15.76 - 16.12 (-1),
16.12 - 16.60 (-3), 16.60 - 18.70 (-0.7), 18.70 - 19.37 (-0.9), 19.37 - 22.37
(-1.67). In the case of quasars we assume 10.5 - 14.83 ($\alpha$=2.5),
14.83 - 15.76 (-0.5), 15.76 - 16.12 (-1), 16.12 - 17.05 (-3), 17.05 - 19.37
(-0.7), 19.37 - 22.37 (-1.67).

\ \\
This paper has been processed by the authors using the Blackwell
Scientific Publications \LaTeX\  style file.

\end{document}